# Self-sorting of Bi-dispersed Colloidal Particles near Contact Line of an Evaporating Sessile Droplet

Nagesh D. Patil, Rajneesh Bhardwaj[*], Atul Sharma

Department of Mechanical Engineering,

Indian Institute of Technology Bombay, Mumbai, 400076 India

[*]Corresponding author (rajneesh.bhardwaj@iitb.ac.in)

Phone: +91 22 2576 7534

*Abstract*

We investigate deposit patterns and associated morphology formed after the evaporation of an aqueous droplet containing mono- and bi-dispersed colloidal particles. In particular, the combined effect of substrate heating and particle diameter is investigated. We employ high-speed visualization, optical microscopy and scanning electron microscopy to characterize the evaporating droplets, particle motion, and deposit morphology, respectively. In the context of mono-dispersed colloidal particles, an inner deposit and a typical ring form for smaller and larger particles, respectively, on a nonheated surface. The formation of the inner deposit is attributed to early depinning of the contact line, explained by a mechanistic model based on the balance of several forces acting on a particle near the contact line. At larger substrate temperature, a thin ring with inner deposit forms, explained by the self-pinning of the contact line and advection of the particles from the contact line to the center of the droplet due to Marangoni flow. In the context of bi-dispersed colloidal particles, self-sorting of the colloidal particles within the ring occurs at larger substrate temperature. The smaller particles deposit at the outermost edge as compared to the larger diameter particles and this preferential deposition in a stagnation region near the contact line is due to the spatially-varying height of the liquid-gas interface above the substrate. The sorting occurs at a smaller ratio of the diameter of the smaller and larger particle. At the larger substrate temperature and a larger ratio, the particles do not get sorted and mix into each other. Our measurements show that there exists a critical substrate temperature as well as a diameter ratio in order to achieve the sorting. We propose regime maps on substrate temperature-particle diameter and substrate temperature-diameter ratio plane for mono- and bi-dispersed solutions, respectively.

## 1. Introduction

In interface science, self-assembly of colloidal particles in an evaporating sessile droplet on a solid substrate has received significant attention recently[1]. Technical applications include manufacturing of colloidal self-assemblies, bioassays, biosensors and ink-jet printing. As explained by Deegan et al.[2], the evaporation mass flux on the liquid-gas interface is non-uniform and is the largest near the contact line for equilibrium contact angle ($\theta_{eq}$) lesser than 90°. The evaporation mass flux together with pinning of the contact line develops a radially outward fluid flow and thereby advecting suspended colloidal particles towards contact line to form a *ring*-like deposit (or coffee-ring effect, Figure 1(a)). Subsequent studies showed effects of thermal Marangoni convection and receding angle on the deposit pattern. If the thermal Marangoni convection is present with a smaller receding angle, the contact line pins because of a stagnation region near the contact line caused by the generation of Marangoni recirculation, outward capillary flow, and spatially-varying height of the liquid-gas interface near the contact line. This results in a *thin ring with inner deposit* pattern, as shown in Figure 1(b). In case of a larger receding angle, the contact line depins early during the evaporation and thereby advecting all particles towards the center of the droplet to form an *inner deposit* in the absence or presence of the Marangoni recirculation[3-4], as shown schematically in Figure 1(c) and 1(d), respectively.

Several previous studies achieved the thermal Marangoni convection in droplets loaded with mono-dispersed colloidal particles by heating the substrate. For instance, Li et al.[5] varied glass substrate temperature ($T_s$ = 30 - 80°C) for the evaporation of droplets containing polystyrene particles of diameter, $d_P$ = 0.1 µm with particles concentration, $c$ = 0.25 % (v/v). The equilibrium contact angles ($\theta_{eq}$) of the droplets were in the range of 24 - 30°. On the nonheated substrate, a ring forms while on the heated substrate a thin ring with inner deposit (called as "coffee eye") forms. Similarly, Parsa et al.[6] observed that a uniform deposit pattern changes to dual-ring and multi-ring deposit patterns with an increase in silicon substrate temperature ($\theta_{eq}$ = 30 - 40°, $T_s$ = 25 - 99°C). They used copper oxide powder having $d_P$ = 50 nm with $c$ = 0.05 % (w/w). Further, Zhong and Duan [7] studied the evaporation of droplets containing graphite nanopowders of $d_P$ = 2-3 nm diameter particles with $c$ = 0.05 % (v/v) on silicon. The substrate temperature was varied from 10°C to 50°C. A uniform deposition of the particles to dual ring (outer thin ring and inner thick ring) deposition of the particles was observed with increase in temperature. Very recently, Patil et al.[8] reported the effect of substrate temperature ($T_s$ = 27 - 90°C) for hydrophilic glass ($\theta_{eq}$ = 34 - 38°) and hydrophobic silicon ($\theta_{eq}$ = 94 - 97°) for evaporation of droplets containing $d_P$ = 0.46 μm polystyrene particles with $c$ = 0.05, 0.1 and 1.0% (v/v). At ambient substrate temperature on glass and silicon a ring and an inner deposit formed, respectively. At larger substrate temperature, a thin ring with inner deposit formed on the glass as well as silicon. In all these studies, the thin ring with inner deposit or the dual ring was attributed to Marangoni convection inside the droplet, that transports the particles from the contact line region to the inner region of the droplet.

The effect of the particle diameter was also investigated for mono-dispersed colloidal particles on a nonheated substrate. Perelaer et al.[9] considered silica particles of $d_P$ = 0.33, 1, 3 and 5 μm ($c$ = 1% w/w) on glass with different wettability ($\theta_{eq}$ = 5, 25, 130°). Biswas et al.[10] considered polystyrene particles of $d_P$ = 0.02, 0.2 and 1.1 μm ($c$ = 0.05% w/w) on glass with different wettability ($\theta_{rec}$ = 6, 17, 89°). The results in both studies showed a ring-like pattern for smaller particle diameters on a smaller contact angle (i.e. hydrophilic) substrates, while an inner deposit pattern for all cases of particle diameter on a larger contact angle (i.e. hydrophobic) substrates. It implies that the particle diameter and substrate wettability are important parameters in determining the final deposit pattern. It was also reported that the particle diameter influences its deposition location in the wedge-shaped region at the contact line. For instance, Weon and Je [11] studied the self-pinning of particles at the contact line by comparing two experiments of different particle diameters, $d_P$ = 0.1 and 1.0 μm ($c$ = 1% w/w) on glass ($\theta_{eq}$ = 15°). They observed that for the same particle concentration after drying of the droplet, the nanoparticles ($d_P$ = 0.1 μm) get closely packed at the contact line and get pinned easily in comparison to the microparticles ($d_P$ = 1.0 μm). Lee et al.[12] studied the deposit patterns of droplets containing $Al_2O_3$ ($d_P$ = 9, 13, 20, 80 and 135 nm) and $TiO_2$ ($d_P$ = 21 nm) nanoparticles for different particle concentrations (0.01-4% v/v) on glass ($\theta_{eq}$ = 55°), stainless steel ($\theta_{eq}$ = 80°) and teflon ($\theta_{eq}$ = 90°). At low concentration (below 2% v/v) and smaller particle diameters ($d_P$ = 9 and 13 nm), ring-like pattern formed on high wettability glass and silicon, while a uniform deposition of particles formed at high concentrations and large particle diameters (20-135 nm) for $Al_2O_3$. Ring-like deposit patterns were observed for all concentrations of $TiO_2$ particles. Recently, Yu et al.[13] investigated the effect of larger polystyrene particles of $d_P$ = 20 μm in water and ethanol/water droplets with $c$ = 0.02-1.28% (w/w) on PDMS substrate ($\theta_{eq}$ = 70-110°). At low concentration of the microparticles, depinning of the contact line occurred and inner deposit with monolayer pattern of particles was obtained, while at high concentration pinning of the contact line occurred and a ring with multilayer pattern of particles was obtained [13]. Authors concluded that the addition of ethanol in the solution does not influence deposit patterns. In order to predict the contact line pinning, they proposed a mechanistic model of self-pinning, based on balance of forces on the particles near the contact line[13].

In the context of bi-dispersed colloidal particles, self-sorting of the particles in the dried deposit was achieved by Han et al.[14] for the bi-dispersed particles of $d_P$ = 50 nm and 500 nm on silicon, with smaller diameter particles deposited closer to the contact line as compared to the larger particles. It was attributed to spatially-varying height of the liquid-gas interface near the contact line. Chhasatia and Sun[15] studied the deposit patterns of an inkjet-printed aqueous colloidal mixture ($d_P$ = 100 nm and 1.1 µm, $c$ = 0.5% v/v each) onto a glass substrate of different wettability ($\theta_{rec}$ = 80, 55, 30, 10, and 0°). They showed that inside evaporating droplet particles rearrange themselves near the droplet contact line according to their diameters; with an increase in the substrate wettability, the particles separation in the deposit patterns was increased. Weon and Je [16] investigated the fingering pattern inside the ring from the bi-dispersed colloidal particles ($d_P$ = 0.2 and 2 µm, $c$ = 0.05% v/v) which were suspended in decalin liquid and were allowed to evaporate on the glass ($\theta_{eq}$ = 20°). At the end of evaporation, the smaller and larger particles assemble into an outer and inner ring, respectively. The fingering of small particles is seen due to the higher Marangoni flow which competes with the outward capillary flow. Devlin et al.[17] obtained the sorting of 1 and 3 µm colloidal particles ($c$ = 0.025% w/w) on a silicon substrate with 1 µm particles forming the outer ring and 3 µm particles inner ring.

The effect of the ratio of the diameter of bi-dispersed colloidal particles ($R$) on a nonheated substrate was studied by Monteux and Lequeux [18] and they reported that at the edge of the droplet a depletion region is created depending on the contact angle ($\theta_{eq}$ = 10°) and particle diameter. This effect was used to sort bi-dispersed particles ($d_P$ = 100 nm and 1 µm ($R$ = 0.1), 100 nm and 5 µm ($R$ = 0.02) and 1 and 5 µm ($R$ = 0.2), $c$ = 0.16-1.6% w/w), where nanometer particles deposit at the edge of the droplet, whereas micrometer particles are restricted further away from the edge.

The effect of the substrate temperature for bi-dispersed colloidal particles ($R$) was studied by Hendarto and Gianchandani [19] and reported the sorting of hollow micrometer glass particles of $d_P$ = 5-200 µm suspended in isopropyl alcohol droplets by varying substrate temperature to $T_s$

= 55-85°C. They attributed the sorting of particles to the Marangoni convection and showed that larger spheres deposited at the center of the droplet while smaller spheres deposited at the outer ring. They also reported that a better sorting is seen at a higher temperature. Very recently, Parsa et al. [20] studied the evaporation of droplets containing bi-dispersed particles of 1 and 3.2 μm diameter on a silicon wafer. They obtained five types of deposit patterns with respect to increase in substrate temperatures ($T_s$ = 22-99°C). At the contact line, they obtained sorting of the particles with 1 μm particles at outermost edge and mixture of 1 and 3.2 μm particles in the inner ring.

To the best of our knowledge, previous studies did not address the combined effect of the substrate temperature and the particle diameter in case of mono- and bi-dispersed colloidal particles. A study by Hendarto and Gianchandani[19] for bi-dispersed particles is an exception, however, in this study, authors used buoyant particles (0.14 gcm$^{-3}$) and the deposits of neutrally buoyant colloidal particles (i.e. focus of the present work) are likely to be different. The second issue in this arena is to predict the deposit pattern as a function of particle diameter and substrate temperature. While there are several studies considered the effects of substrate temperature, particle diameter, particles concentration, and substrate wettability, consolidated regime maps to predict the deposit type are scarcely available in the literature. Therefore, the objective of this work is to study the combined effect of substrate heating and particle diameter on the deposit patterns obtained from the evaporation of mono-and bi-dispersed colloidal particles suspended in aqueous droplets on a silicon wafer. A secondary objective is to propose a regime map as a function of substrate temperature and particle diameter in order to predict the deposit pattern. We extend our previous study[8] in which we investigated the effect of substrate heating and wettability on deposit patterns during evaporation of droplet containing 0.46 μm neutrally buoyant polystyrene particles on the hydrophilic glass and the hydrophobic wet-oxidized silicon wafer. In the present study, we consider neutrally buoyant polystyrene particles (spheres having a density of 1.05 gcm$^{-3}$) in an aqueous droplet evaporating on a wet-oxidized silicon wafer. We vary particle diameter ($d_P$ = 0.1, 0.46, 1.1 and 3 μm) for mono-dispersed particles solution, and particle diameter ratio ($R$ = 0.03, 0.09, 0.15, 0.21 and 0.36) for the bi-dispersed particles solution.

## 2. Methods

### 2.1 Experimental details

Aqueous colloidal suspensions of 10% concentration ($c$, v/v) of mono-dispersed polystyrene latex beads of different diameters were obtained from Sigma Aldrich Inc, USA. The following diameters were selected: 0.1, 0.46, 1.1, and 3 μm. The solutions were diluted to $c = 0.05\%$, as discussed in our earlier work[8]. The bi-dispersed solutions are made by mixing an equal volume of solutions of two different particle diameters. The solutions were stabilized with sufficient sonication (about 1 hour) and ensured that no sedimentation or agglomeration of particles occurred. The following particle diameter combinations are used: 0.1 and 3 μm (particle diameter ratio, $R = 0.03$), 0.1 and 1.1 μm (0.09), 0.46 and 3 μm (0.15), 0.1 and 0.46 μm (0.21), and 1.1 and 3 μm (0.36) were prepared.

Droplets of 1.1±0.2 μL volume were generated using a micropipette (Prime, Biosystem Diagnostics Inc., India) and were gently deposited on a silicon wafer (diameter = 51 mm, thickness = 0.3 mm). The polished side of the wafer was cleaned by standard RCA cleaning process and later wet oxidized in a furnace. Time-varying evaporating droplet shapes from the side were recorded by using a high-speed camera. The schematic of the setup and other relevant details can be found in our previous work[8]. The images were recorded at 10 and 100 frames per second for nonheated and heated substrates, respectively. The experiments were performed at ambient ($T_s = 27°C$) and two cases of substrate temperatures ($T_s = 60$ and $90°C$). The wafer was heated from the bottom by a digital hot plate. The equilibrium ($\theta_{eq}$) and receding ($\theta_{rec}$) contact angles measured on the surface for sessile droplets of mono-dispersed as well as bi-dispersed solutions for nonheated/heated silicon are listed in Table 1 and Table 2, respectively. The uncertainty in measurements of wetted diameter is around ±28 μm. Each set of experiment was performed three times. The values of ambient temperature and relative humidity were 27.0±1.5°C and 35±5%, respectively. After droplet evaporation, the deposit patterns images were taken by an optical microscope (Olympus MX-40, objective lens 2.5X to 50X). The morphology of the deposits was characterized by FEG-SEM (field emission gun scanning electron microscopy, Model- JSM-7600F, JEOL Ltd.) with a sub-micrometer resolution.

### 2.2 A theoretical model of the depinning of the contact line

As explained by Jung et al.[21] and Wong et al.[22], during the evaporation of a droplet containing colloidal particles, particles at the contact line experience several forces as shown by a schematic in Figure 2[13]. As shown in the figure, the surface tension force ($F_s$) acting on the particles has a tendency to pull the particles away from the contact line while the friction force (due to the adhesion forces) has a tendency to keep the particles stationary. The total adhesion force ($F_a$) acting between particles and the substrate is the summation of the van der Waals force ($F_{wps}$) between particles and the substrate in a fluid medium, and electrostatic forces ($F_{eps}$), given by eq. (1) and (2), respectively[15, 23]. Here, the gravitational force can be neglected as the density of polystyrene particle is almost same as that of water.

$$F_{wps} = \frac{2A_{123}(d_P/2)^3}{3z^2(z+d_P)^2} \quad (1)$$

$$F_{eps} = -\pi d_P \varepsilon k \frac{\left[\phi_1^2 + \phi_2^2 - 2\phi_1\phi_2 \exp(kz)\right]}{\left[\exp(2kz)-1\right]} \quad (2)$$

The symbols in eqs. 1-2 are defined in the supporting information. At the contact line, the outermost particle (see Figure 2) experiences a surface tension force, $F_s$, due to the liquid-gas interface given by Eq. (4) [13, 24-25]. The liquid-gas interface touching this particle is shown in Figure 2. The surface tension force at a point is given as $\gamma_{LG} \cos\phi$ and $\gamma_{LG} \sin\phi$, where $\gamma_{LG}$ is the surface tension of water (i.e. liquid-vapor interface) and $\pi - 2\phi$ is the angle of liquid layer wetting the outermost particle, where $\phi$ is an angle shown in Figure 2.

$$F_s = \int dF_s = \int \gamma_{LG} \cos\phi \, dr \quad (3)$$

$$F_s = \pi d_P \gamma_{LG} \cos\phi \quad (4)$$

The vertical component of the surface tension force will be $F_s \cos\theta_{rec}$ which acts as an adhering force and horizontal or parallel component of this force will be $F_s \sin\theta_{rec}$ which acts as a pulling force inside the liquid in radial direction [25]. The total adhesion force ($F_{a1}$) acting on the outermost particle is given by [13, 25],

$$F_{a1} = F_{wps1} + F_{eps1} + F_s \cos\theta_{rec} \quad (5)$$

where $\theta_{rec}$ is the receding contact angle of the liquid-gas interface. The adhesion force acting on the inner particle near the outermost particle is given by $F_{a2} = F_{wps2} + F_{eps2}$. Thus, at the contact line balancing all the forces acting on the particles parallel to the substrate, a self-pinning mechanism for the particles at the contact line is recently given by Yu et al.[13, 25] as follows,

$$\Delta F = F_s \sin\theta - \left[ f\left(F_s \cos\theta + nF_a\right) + nF_d \right] \leq 0 \qquad (6)$$

where $f$ is a friction coefficient between the particles and the substrate in a liquid medium, $n$ is the number of the particles required for inducing early pinning of the contact line. $F_d = 3\pi d_P \mu v_{rad}$ is the drag force acting on the particles [21, 26], $\mu$ is the dynamic viscosity of the water, $v_{rad}$ is the velocity of the evaporative driven flow, $(v_{rad} \sim J_{max}/\rho_L)$, $J_{max}$ is the maximum evaporative flux near the contact line and $\rho_L$ is droplet density [26-27]. In Figure 2, $fF_a$ is the friction force acting on the particles. The resultant force $\Delta F$ is non-dimensionalized as follows,

$$\Delta F^* = \frac{\Delta F}{\gamma_{LG} d_p} \qquad (7)$$

All parameter values required for the calculation of the different forces are taken from previous studies[22, 28-30] and are given in supporting information.

## 3. Results and discussions

In the following subsections, we discuss the combined effect of the substrate temperature and particle size on the deposit pattern and morphology of mono-dispersed and bi-dispersed colloidal particles.

### 3.1 Mono-dispersed colloidal particles

#### 3.1.1 Measurements of deposit patterns

The different deposit shapes for different particle diameters ($d_p$ = 0.1, 0.46, 1.1 and 3 µm) and substrate temperatures ($T_s$ = 27, 60 and 90°C) are shown in Figure 3. The frames in rows and columns show the deposits at constant particle diameter and at constant substrate temperature, respectively. Zoomed-in view of a part of the deposit recorded by the SEM is also shown for some cases. The images obtained in two different runs for each experiment are plotted in supporting information (Figs. S1, S2 and S3). The datasets obtained in the present measurements show good repeatability. Figure 3(a) shows the deposit patterns for 0.1, 0.46, 1.1 and 3 µm particles at $T_s$ = 27°C (nonheated case). The ring does not form for smaller particles ($d_p$ = 0.1, 0.46 and 1.1 µm) due to early depinning of the contact line, and particles move inside towards the center of the droplet and finally form the inner deposit, with a faint appearance of a ring in these cases. The inner deposit formed for 0.46 µm case is consistent with our previous study[8]. By contrast, a ring forms for $d_p$ = 3 µm due to pinning of the contact line to the substrate. In this case, the particles are drawn to the contact line due to an evaporative driven radial outward flow to form a ring-like deposit[2].

The time-varying non-dimensional droplet wetted diameter ($D_{wetted}$) and contact angle ($\theta$) are plotted in Figure 4. $D_{wetted}$ and $\theta$ are normalized with respect to initial wetted diameter and equilibrium contact angle. For $d_p$ = 0.1, 0.46 and 1.1 µm, initially for $t < 200$ s), $D_{wetted}$ is constant and $\theta$ decreases with time. Further, after $t = 200$ s, the droplet attains its receding contact angle and $D_{wetted}$ decreases with time due to depinning of the contact line until the complete evaporation. For $d_p = 3$ µm, during the complete evaporation of droplet $D_{wetted}$ remains constant (i.e. contact line pins) and $\theta$ decreases with time (Figure 4), exhibiting the pinning of the contact line. The measured receding contact angles for smaller particles are high as compared to the larger particle, as listed in Table 1. In Figure 4, the evaporation for $d_p = 3$ µm is faster than other cases ($d_p \leq 1.1$ µm) since the pinning in the former increases surface area of the droplet.

In Figure 3(b) and (c), deposit patterns are shown at two substrate temperatures $T_s$ = 60 and 90°C, respectively, for all cases of particle diameter. It is found that a ring with inner deposits forms for all cases similar to the deposit patterns obtained in our previous work [8] for 0.46 µm. Due to the substrate heating, the thermal Marangoni flow develops from contact line region towards the droplet apex-point (Figure 1(b)). Consequently, the particles are carried towards contact line by radial outward flow and the circulating Marangoni flow advects the particles away from the contact line, as shown by schematic in Figure 1(b). As explained in previous studies [5, 8], a stagnation-region develops near the contact line region and some particles get trapped in this region and aid to the pinning of the contact line (Figure 1(b)). Thus, even smaller particles ($d_p$ = 0.1, 0.46 and 1.1 µm) on a heated substrate ($T_s$ = 60 and 90°C) offer resistance against the depinning. Because of the pinning, some particles deposit in this region, while most of the particles are deposited as an inner deposit due to a stronger Marangoni flow. The magnitude of Marangoni velocity scales with the temperature gradient across the interface [26, 31] and the temperature gradient increases with substrate temperature. Therefore, *a thin ring with inner deposit* forms due to the combined effect of particle size and substrate heating, as shown in Figure 3(b) and (c). Table 1 lists measured receding contact angles for all cases of substrate heating. The receding contact angles are very smaller and are on the same order. It implies that the most of the evaporation occurs in pinned contact line mode leading to the ring with inner deposits.

SEM images plotted in Figure 3(c) qualitatively shows that the thinning of the ring width occurs with an increase in substrate temperature (second row for $d_p$ = 1.1 and 3 μm).. A quantitative measurement of ring profile for each case is presented in the supporting information (Figure S4). From the quantitative data, we observe significant thinning of the ring width for $d_P$ = 1.1 and 3 μm while for $d_P$ = 0.1 and 0.46 μm, the ring width is almost same. This is attributed to an increase in the magnitude of Marangoni convection for higher temperature which carries more particles towards the inner deposit[8]. In other words, due to the stronger Marangoni convection at $T_s$ = 90°C, size of stagnation region is lesser[5] and thereby, few particles deposit in the ring and more particles deposit in the inner region.

### 3.1.2 Comparison between measurements and predictions of model of depinning of the contact line

We verify the self-pinning by larger particles described in section 3.1.1 by a model described in section 2.2. In this model, we assume that the deposited particles are stacked in a monolayer near the contact line, valid for low particles concentration. The model proposed by Yu et al.[13, 25] considers several forces on particles near the contact line and if the horizontal force in an outward direction (friction and drag force) exceeds that in the inward direction (surface tension force), the pinning occurs. We plot $\Delta F^*$ (eq. 7) in Figure 5 against the diameter of the particle. We test the model for two values of the coefficient of friction, $f$ = 0.1 and 0.9. We also vary number of particles near the contact line, $n$ = 1 and 12. The maximum value of $n$, $n$ = 12 was estimated from measurements. In Figure 3, at $T_S$ = 27°C= and $d_P$ = 3 μm, the contact line pins and ring forms with around 12-14 particles in it (first column, last row). In order to test the sensitivity of $\phi$ (as defined in Figure 2), we considered the lowest value $\phi$ = 1° and the highest value $\phi$ = 89° (closer to the critical angle of wetting layer[13, 25]) for the analysis. The former value implies that the half of the surface area of the particle is dewetted while the latter value indicates that liquid-gas interface is touching at a point. In Figure 5(a), for $n$ = 1 and $\phi$ = 1°, we calculated the resultant force; for all particle diameters, the values are positive indicating that the surface tension force $F_s \sin\theta_{rec}$ dominates the resistance force $\left[ f\left(F_s \cos\theta_{rec} + nF_a\right) + nF_d \right]$ and depinning of the contact line occurs. Further, for $n$ =12 and $\phi$ = 1°, the values for $d_p$ = 0.1, 0.46 and 1.1 μm are positive and for $d_p$ = 3 μm is negative value. A negative value indicates that the resistance force $\left[ f\left( F_s \cos\theta_{rec} + nF_a \right) + nF_d \right]$ is larger than driving surface tension force $F_s \sin\theta_{rec}$. Here the number of particles $n$ are increased which increases the friction force and thereby the self-pinning of $d_p = 3$ µm occurs (as noted in Figure 3a, fourth row). Other particle diameters do not experience the sufficient friction force near the contact line in order to get pinned. They are overcome by surface tension force $F_s \sin\theta_{rec}$ and the particles get pulled inwards, and inner deposit forms (as can be seen in Figure 3a, first, second, the third column). For $n = 1$ and $\phi = 89°$, for $d_p = 0.1$, 0.46 and 1.1 µm the depinning of contact line is predicted and for $d_p = 3$ µm the pinning of contact line is predicted. Thus, it indicates less wetting on the particle offers more adhesion force to the substrate and pinning increases. For $n = 12$ and $\phi = 89°$, for all particle diameters, the pinning of contact line is predicted as the number of particles are increased, offering more resistance to overcome the pulling surface tension force. We also tested this model for a higher coefficient of friction $f = 0.9$ (see Figure 5b), for cases of $n = 1$ and 12 and $\phi = 1°$ and 89°. It is observed that with an increase in friction coefficient the pinning of particles at the contact line increases.

### 3.1.3 Regime map

We propose a regime map in Figure 6 as a function of substrate temperature and diameter of particles for predicting the deposit pattern of mono-dispersed particles. The map shows three regimes, namely, *inner deposit*, *ring*, and *thin ring with inner deposit*. The insets show the optical microscopic (full view) and SEM images (zoomed view of the part of the ring) of the deposits. The first regime corresponds to *inner deposit* that forms at smaller particle diameter and lower substrate temperature. A representative image of the deposit is shown in an inset (bottom left) of Figure 6. As explained earlier, this deposit pattern forms due to the early receding of the contact line. The depinning occurs because the surface tension force overcomes lower friction force between the particle and substrate (section 3.1.2). Thus, all particles are advected inwards to form an inner deposit.

The second regime corresponds to *ring* which forms at larger particle diameter and low substrate temperature. A representative image of the deposit is shown in an inset (top left) of Figure 6. As the particle diameter is larger, the larger friction force overcomes inward pulling surface tension force and the pinning of particles at the contact line occurs (section 3.1.2). Owing to the radially outward flow due to the maximum evaporation near the contact line, the particles get advected to the contact line and the ring forms.

The third regime corresponds to *a thin ring with inner deposit* which forms at a larger substrate temperature for all particle diameters. This deposit pattern is formed due to the strong Marangoni flow inside the droplet which advects most of the particles in the inner region of the droplet and remaining particles are deposited at the contact line due to the formation of a stagnation region [5, 8]. The ring becomes thinner at a larger substrate temperature because the intensity of the Marangoni convection increases with the substrate temperature [8]. Therefore, at a lower substrate temperature, as the particle diameter increases from smaller to larger, the regime changes from the inner deposit to the ring; and as the substrate temperature increases, both these regimes changes to the thin ring with inner deposit.

## 3.2 Bi-dispersed colloidal particles

### 3.2.1 Measurements of deposit patterns

In this section, the combined effect of substrate heating and particle diameter ratio on the deposit patterns from the evaporation of droplets containing bi-dispersed colloidal particles on a wet-oxidized silicon wafer are studied. We considered five particle diameter ratios, $R$, defined as the ratio of the diameter of the smaller particle to the diameter of the larger particle. In the present measurements, several cases of $R$ and substrate temperatures are taken as follows: $R = 0.03, 0.09, 0.15, 0.21$ and $0.36$ and $T_s = 27, 60$ and $90°C$. In Figure 7 (first row), the deposit patterns obtained for $R = 0.03$ ($d_P = 0.1$ and $3$ µm) are shown, in which an inner deposit forms at $T_s = 27°C$ and a thin ring with inner deposit forms at $T_s = 60$ and $90°C$. Note that there a faint appearance of a ring in the former case. Insets in Figure 7 (second and third row) show the deposit patterns (SEM images) at two resolution levels. The datasets obtained in two different runs for each experiment are plotted in supporting information (Figs. S5, S6 and S7) and demonstrate good repeatability in the measurements. The inner deposit forms owing to the receding of contact line and the thin ring with inner deposit forms due to the pinning of contact line aided by the substrate heating, as explained in the previous section as well as by Patil et al.[8]. The values of the measured receding contact angle for all cases are given in Table 2.

We record patterns of particles of two different diameters near the contact line in Figure 7 (second row). At $T_s = 27°C$, 0.1 µm, and 3 µm particles are mixed while at $T_s = 60$ and $90°C$, 0.1 µm particles are deposited at the outer edge as compared to 3 µm particles. Due to substrate heating, a temperature gradient exists along the liquid-gas interface which generates the circulating Marangoni flow, evidenced in our earlier work [8]. This results in formation of a thin ring with an inner deposit [8]. On the heated substrate cases, the segregation or sorting of the particles in the thin ring occurs in the stagnation region. The formation of the stagnation region was reported in previous studies [5, 8]. The particles are trapped and self-sorted in this region according to their diameters, as shown in a schematic in Figure 8(a). A similar hypothesis was proposed in previous studies [9-10, 15, 22]. Therefore, the smaller particles deposit very near to the contact line as compared to the larger particles, as shown in a schematic in Figure 8(b). In Figure 7 (third row), the images are shown with a larger magnification, that distinctively shows that at the outermost edge 0.1 µm particles are deposited while a mixture of 0.1 and 3 µm diameter particles is seen inside adjacent to the deposits of 0.1 µm particles.

The deposit patterns for larger particle diameter ratio, $R = 0.09$ ($d_P = 0.1$ and 1.1 µm) and 0.15 ($d_P = 0.46$ and 3 µm) are shown in Figure 9 and Figure 10, respectively. At $T_s = 27°C$, the inner deposit forms due to the depinning of the contact line for both cases of diameter ratio ($R$). The deposit is a mixture of 0.1 µm and 1.1 µm particles (Figure 9, first column), and 0.46 and 3 µm particles (Figure 10, first column). At $T_s = 60$ and 90°C, the thin ring with inner deposit forms for both particle diameter ratio, similar to that observed for $R = 0.03$. The zoomed view of the morphology of the deposit is evidenced in the second row of Figure 9 and Figure 10.

In Figure 11 and Figure 12, we further tested a larger $R$, $R = 0.21$ ($d_P = 0.1$ and 0.46 µm) and $R = 0.36$ ($d_P = 1.1$ and 3 µm), respectively. At $T_s = 27°C$, the inner deposit forms for both cases with a mixture of colloidal particles. At $T_s = 60$ and 90°C, a thin ring with inner deposit forms. Note that the deposit is a mixture of the colloidal particles in the ring. This is due to comparable diameters of the particles, as shown in the schematic of Figure 13(a). The self-sorting of the particles does not occur at the contact line. A schematic of the mechanism of *no self-sorting* within the formed ring is shown in Figure 13(b). The diameter 0.1 and 0.46 µm particles are on same order, therefore both diameter particles are penetrated uniformly in the contact line region and sorting is not observed, as shown in Figure 11 (third row). Similarly, in case of 1.1 and 3 µm colloidal mixture ($R = 0.36$), the self-sorting of the particles at the contact line is not achieved (Figure 12, second and third column).

### 3.2.2 Comparison between measurements and predictions of the model

The obtained deposit patterns with a sorting or no-sorting of particles in the ring for bi-dispersed particles can be explained by a simple geometric model, similar to the model reported by Perelaer et al.[9] and Biswas et al.[10]. During the evaporation on a heated substrate, the depositing particles near the contact line penetrate in the wedge-shaped region of the contact line, as shown in a schematic in Figure 8(a). In this region, the location of the deposition of a particle depends upon its diameter and the smaller diameter particles can be easily driven closer to the contact line than the larger diameter particles (Figure 8(a)). The schematic in Figure 8(a) shows two different diameter particles with diameters $d_{P1}$ and $d_{P2}$ such that $d_{P1} > d_{P2}$. The separation distance ($x_{theoretical}$) between these two particles is given by [9-10],

$$x_{theoretical} = \frac{d_{P1} - d_{P2}}{2\tan(\theta_{eq}/2)} \quad (8)$$

where $\theta_{eq}$ is the equilibrium contact angle. We compare predictions of this simple model with our measurements for three particle diameter ratios $R$ = 0.03, 0.09 and 0.15, at which the self-sorting occurs. The theoretical values, $x_{theoretical}$, obtained by the model (eq. 8) are 1.61, 0.55 and 1.4 µm, respectively and the measured values (at $T_s$ = 90°C) are 1.72, 0.6, and 2.2 µm, respectively. Further, we compare the results for $R$ = 0.21 ($d_P$ = 0.1 and 0.46 µm, Figure 11) and 0.36 ($d_P$ = 1.1 and 3 µm, Figure 12), at which sorting does not occur. For these particle diameter ratios, the theoretical values, $x_{theoretical}$, obtained using eq. 8 are 0.19 and 1.05 µm, respectively. The theoretical values are on the same order as the respective smaller particle diameter in the bi-dispersed colloidal solution. Therefore, particles of both diameters deposit closer to each other as shown in a schematic in Figure 13(a) and a mixture of the deposits forms without sorting.

### 3.2.3 Regime map

We propose a regime map for the bi-dispersed colloidal deposits in Figure 14, as a function of the substrate temperature and the particle diameter ratio. The map shows three regimes namely, *mixed particles in inner deposit*, *self-sorting in ring* and *no self-sorting in ring*, presented in the following paragraphs. The three regimes are shown with the demarcation of black dotted lines. Insets of the figure show representative SEM images of the deposits for all regimes.

The first regime corresponds to the *mixed particles in the inner deposit* which forms in all cases of particle diameter ratio and at low substrate temperature. The inner deposit is a mixture of the bi-dispersed particles. The representative images of the deposit are shown by insets at the bottom left (low particle diameter ratio) and top left (high particle diameter ratio). As explained earlier, this deposit pattern forms due to early depinning of the contact line and thereby, the particles follow the motion of contact line and are dragged inside the droplet.

The second regime corresponds to *self-sorting in ring* which forms at small particle diameter ratio and high substrate temperature. A representative image of this deposit (full ring and zoomed view of the part of the ring) is shown by an inset at the bottom right of Figure 14. This deposit pattern forms owing to the pinning of contact line due to the substrate heating and formation of the wedge-shaped stagnation region near the contact line. As explained earlier, if the ratio of the diameter of the particles ($R$) is smaller, the self-sorting occurs at the contact line, with deposition of the smaller diameter particles at the outermost edge as compared to the larger diameter particles.

The third regime corresponds to *no self-sorting in the ring*, which forms at larger particle diameter ratio and higher substrate temperature. The deposit in the ring is a mixture of the bi-dispersed particles, without sorting. A representative image of this deposit (a full ring and zoomed view of the part of the ring) is shown by an inset at the top right of Figure 14. This deposit pattern forms owing to the pinning of contact line due to the substrate heating. In this case, diameters of the smaller and the larger particle in the bi-dispersed mixture are almost comparable and separation distance between two particles (eq. 8) is on the same order of the smaller diameter. Therefore, no self-sorting of particles at the contact line is evidenced. This also indicates that a critical particle diameter ratio and substrate temperature exist in order to self-sort the particles at the contact line.

## 4. Conclusions

The deposit pattern and morphology formed after the evaporation of a sessile water droplet containing mono- and bi-dispersed colloidal particles have been experimentally investigated. In particular, the combined effect of the substrate heating and particle diameter have been investigated. The following experimental methods are combined to collect the data: high-speed visualization, optical microscopy and scanning electron microscopy. For mono-dispersed particles, with an increase in particle diameters the possibility of pinning of the contact line increases. Therefore, at ambient substrate temperature on the wet-oxidized silicon surface, smaller diameter particles give inner deposits while larger diameter particles give ring formation.

The results are consistent with a model of the depinning of contact line reported in the literature. On heated silicon, for all particle diameters and all temperature, a thin ring with inner deposit forms. The receding contact angle reduces due to the substrate heating and it helps to pin the contact line to form a thin ring, and the developed Marangoni flow inside the droplet advects most of the particles to form the inner deposit. In addition, with an increase in the substrate temperature ring width decreases due to a stronger Marangoni flow inside the droplet, which advects more particles inside the droplet. A regime map is proposed to predict the deposit patterns which shows three deposits namely, *inner deposit*, *ring*, and *thin ring with inner deposit*.

In the context of bi-dispersed particles, the effect of particle-diameter ratio and the substrate temperature is investigated. On nonheated silicon, an inner deposit occurs due to the early depinning of the contact line; while on heated silicon, self-sorting of the particles at the contact line is observed within the ring at the contact line. SEM images reveal such deposit patterns of bi-dispersed colloidal particles. The self-sorting within the ring occurs due to the spatially-varying height of the liquid-gas interface near the contact line. The smaller diameter particles deposit at the outermost edge as compared to the larger diameter particles, which deposit adjacent to the smaller ones. At the contact line due to the substrate heating, a stagnation region is developed by the Marangoni flow and the receding contact angle reduces. For larger particle-diameter ratio, the self-sorting of particles at the contact line is not found due to the comparable diameters of the particles in the mixture. Measurements show that there exists a critical particle-diameter ratio for the sorting of particles. A regime map is proposed for the deposit patterns based on the particle diameter ratio tested here which shows three deposits namely, *mixed particles in inner deposit*, *self-sorting in ring* and *no self-sorting in ring*. The present results provide fundamental insights into the deposit patterns of mono- and bi-dispersed particles and could be useful to control the self-assemblies of particles during the evaporation.

## 5. Supporting Information

Parameter values used in calculations of forces acting on the particles, Deposit patterns for mono-dispersed colloidal particles obtained in two different runs for each experiment, Quantitative measurements of variation of ring profile, Deposit patterns for bi-dispersed colloidal particles obtained in two different runs for each experiment (PDF).

## 6. Acknowledgements

R.B. gratefully acknowledges the financial support of a CSR grant from Portescap Inc. India and of an internal grant from Industrial Research and Consultancy Centre (IRCC), IIT Bombay. N.D.P. was supported by a Ph.D. fellowship awarded by IRCC. The SEM images were recorded at Sophisticated Analytical Instrument Facility (SAIF), IIT Bombay. We also thank anonymous reviewers for their useful comments.

## 7. References

1. Larson, R. G., In Retrospect: Twenty years of drying droplets. *Nature* **2017,** *550* (7677), 466-467.
2. Deegan, R. D.; Bakajin, O.; Dupont, T. F.; Huber, G.; Nagel, S. R.; Witten, T. A., Capillary flow as the cause of ring stains from dried liquid drops. *Nature* **1997,** *389*, 827-829.
3. Hu, H.; Larson, R. G., Marangoni effect reverses coffee-ring depositions. *The Journal of Physical Chemistry B* **2006,** *110* (14), 7090-7094.
4. Bhardwaj, R.; Fang, X.; Attinger, D., Pattern formation during the evaporation of a colloidal nanoliter drop: a numerical and experimental study. *New Journal of Physics* **2009,** *11*, 075020.
5. Li, Y.; Lv, C.; Li, Z.; Quéré, D.; Zheng, Q., From coffee rings to coffee eyes. *Soft matter* **2015,** *11* (23), 4669−4673.
6. Parsa, M.; Harmand, S.; Sefiane, K.; Bigerelle, M.; Deltombe, R., Effect of Substrate Temperature on Pattern Formation of Nanoparticles from Volatile Drops. *Langmuir* **2015,** *31* (11), 3354-3367.
7. Zhong, X.; Duan, F., Disk to dual ring deposition transformation in evaporating nanofluid droplets from substrate cooling to heating. *Physical Chemistry Chemical Physics* **2016,** *18* (30), 20664-20671.
8. Patil, N. D.; Bange, P. G.; Bhardwaj, R.; Sharma, A., Effects of Substrate Heating and Wettability on Evaporation Dynamics and Deposition Patterns for a Sessile Water Droplet Containing Colloidal Particles. *Langmuir* **2016,** *32* (45), 11958-11972.
9. Perelaer, J.; Smith, P. J.; Hendriks, C. E.; van den Berg, A. M.; Schubert, U. S., The preferential deposition of silica micro-particles at the boundary of inkjet printed droplets. *Soft Matter* **2008,** *4* (5), 1072-1078.
10. Biswas, S.; Gawande, S.; Bromberg, V.; Sun, Y., Effects of particle size and substrate surface properties on deposition dynamics of inkjet-printed colloidal drops for printable photovoltaics fabrication. *Journal of Solar Energy Engineering* **2010,** *132* (2), 021010.
11. Weon, B. M.; Je, J. H., Self-pinning by colloids confined at a contact line. *Physical review letters* **2013,** *110* (2), 028303.
12. Lee, H. H.; Fu, S. C.; Tso, C. Y.; Chao, C. Y., Study of residue patterns of aqueous nanofluid droplets with different particle sizes and concentrations on different substrates. *International Journal of Heat and Mass Transfer* **2017,** *105*, 230-236.
13. Yu, Y.-S.; Wang, M.-C.; Huang, X., Evaporative deposition of polystyrene microparticles on PDMS surface. *Scientific Reports* **2017,** *7* (1), 14118.
14. Han, W.; Byun, M.; Lin, Z., Assembling and positioning latex nanoparticles via controlled evaporative self-assembly. *Journal of Materials Chemistry* **2011,** *21* (42), 16968-16972.
15. Chhasatia, V. H.; Sun, Y., Interaction of bi-dispersed particles with contact line in an evaporating colloidal drop. *Soft Matter* **2011,** *7* (21), 10135-10143.
16. Weon, B. M.; Je, J. H., Fingering inside the coffee ring. *Physical Review E* **2013,** *87* (1), 013003.
17. Devlin, N. R.; Loehr, K.; Harris, M. T., The separation of two different sized particles in an evaporating droplet. *AIChE Journal* **2015,** *61* (10), 3547-3556.
18. Monteux, C. c.; Lequeux, F. o., Packing and sorting colloids at the contact line of a drying drop. *Langmuir* **2011,** *27* (6), 2917-2922.
19. Hendarto, E.; Gianchandani, Y. B., Size sorting of floating spheres based on Marangoni forces in evaporating droplets. *Journal of Micromechanics and Microengineering* **2013,** *23* (7), 075016.
20. Parsa, M.; Harmand, S.; Sefiane, K.; Bigerelle, M.; Deltombe, R., Effect of Substrate Temperature on Pattern Formation of Bi-Dispersed Particles from Volatile Drops. *The Journal of Physical Chemistry B* **2017,** *121* (48), 11002-11017.

21. Jung, J.-y.; Kim, Y. W.; Yoo, J. Y.; Koo, J.; Kang, Y. T., Forces acting on a single particle in an evaporating sessile droplet on a hydrophilic surface. *Analytical chemistry* **2010,** *82* (3), 784-788.
22. Wong, T.-S.; Chen, T.-H.; Shen, X.; Ho, C.-M., Nanochromatography driven by the coffee ring effect. *Analytical chemistry* **2011,** *83* (6), 1871-1873.
23. Elimelech, M.; Gregory, J.; Jia, X.; Williams, R. A., *Particle Deposition & Aggregation*. Butterworth Heinemann publications: Woburn, MA, 1998.
24. Sangani, A. S.; Lu, C.; Su, K.; Schwarz, J. A., Capillary force on particles near a drop edge resting on a substrate and a criterion for contact line pinning. *Physical Review E* **2009,** *80* (1), 011603.
25. Yu, Y.-S.; Xia, X.-L.; Zheng, X.; Huang, X.; Zhou, J.-Z., Quasi-static motion of microparticles at the depinning contact line of an evaporating droplet on PDMS surface. *Science China- Physics, Mechanics & Astronomy* **2017,** *60* (9), 094612.
26. Bhardwaj, R.; Fang, X.; Somasundaran, P.; Attinger, D., Self-assembly of colloidal particles from evaporating droplets: role of DLVO interactions and proposition of a phase diagram. *Langmuir* **2010,** *26* (11), 7833-7842.
27. Hu, H.; Larson, R. G., Evaporation of a sessile droplet on a substrate. *J. Phys. Chem. B* **2002,** *106*, 1334-1344.
28. Visser, J., The concept of negative Hamaker coefficients. 1. History and present status. *Advances in Colloid and Interface Science* **1981,** *15* (2), 157-169.
29. Zhang, F.; Busnaina, A. A.; Fury, M. A.; Wang, S.-Q., The removal of deformed submicron particles from silicon wafers by spin rinse and megasonics. *Journal of Electronic Materials* **2000,** *29* (2), 199-204.
30. Qin, K.; Li, Y., Mechanisms of particle removal from silicon wafer surface in wet chemical cleaning process. *Journal of colloid and interface science* **2003,** *261* (2), 569-574.

## 8. Tables

Table 1: Measured receding ($\theta_{rec}$) contact angles during evaporation of droplets with mono-dispersed colloidal particles. The equilibrium contact angle is around 85±2° in all cases.

| $d_P$ (μm) | $\theta_{rec}$ (degrees) | | |
|---|---|---|---|
| | $T_s = 27°C$ | $T_s = 60°C$ | $T_s = 90°C$ |
| 0.1 | 67.8 | 6.2 | 5.7 |
| 0.46 | 67.4 | 4.5 | 6.5 |
| 1.1 | 65.5 | 5.7 | 4.3 |
| 3 | 6.8 | 3.2 | 3.8 |

Table 2: Measured receding ($\theta_{rec}$) contact angles during evaporation of droplets with bi-dispersed colloidal particles. The equilibrium contact angle is around 85±2° in all cases.

| $d_P$ (μm) | Ratio (R) | $\theta_{rec}$ (degrees) | | |
|---|---|---|---|---|
| | | $T_s = 27°C$ | $T_s = 60°C$ | $T_s = 90°C$ |
| 0.1 and 3 | 0.03 | 65.2 | 5.1 | 6.7 |
| 0.1 and 1.1 | 0.09 | 55.3 | 4.2 | 5.2 |
| 0.46 and 3 | 0.15 | 61.9 | 4.7 | 5.6 |
| 0.1 and 0.46 | 0.21 | 60.5 | 5.8 | 6.4 |
| 1.1 and 3 | 0.36 | 60.6 | 3.6 | 3.7 |

## 9. Figures

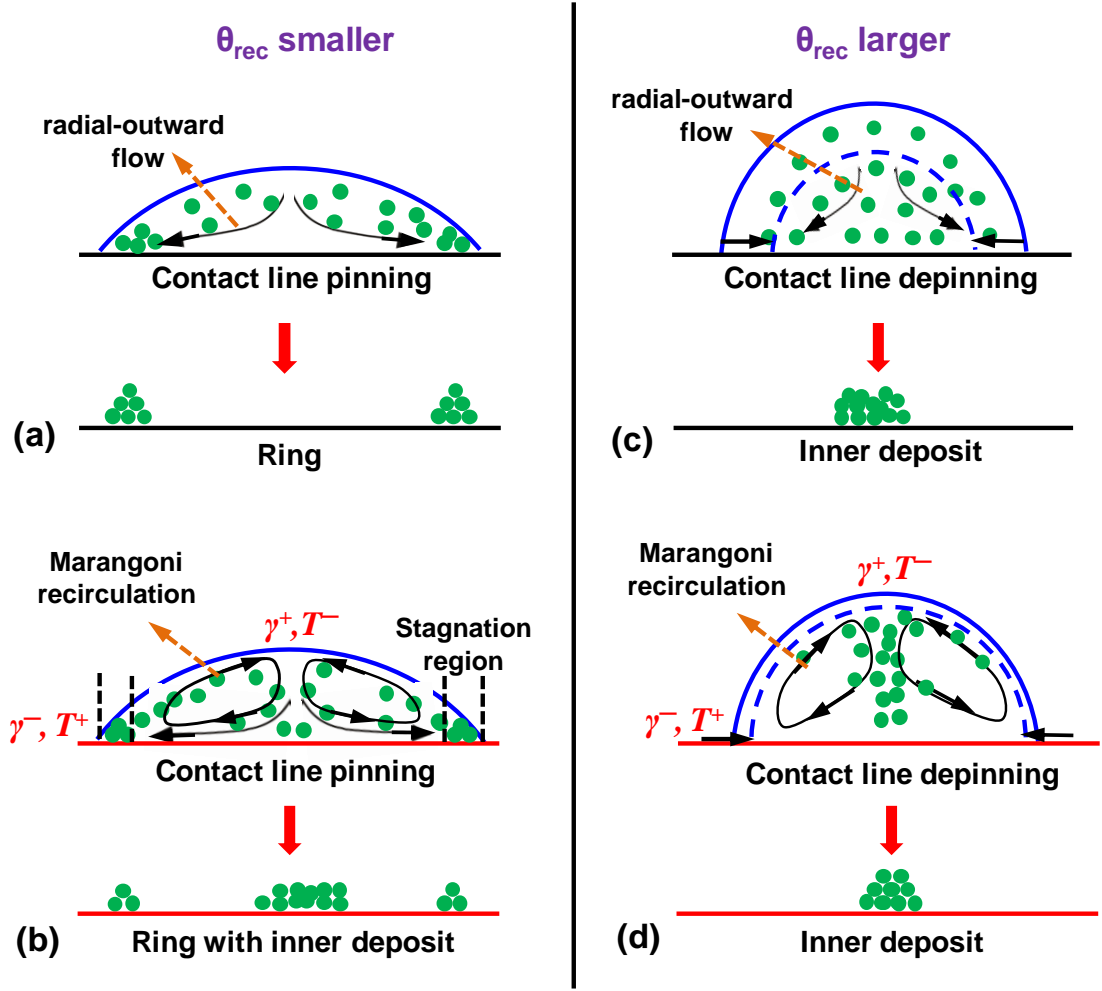

Figure 1: Schematic of the evaporation mechanisms and particle deposit patterns formation on the substrate having (a, b) smaller, and (c, d) larger receding contact angles ($\theta_{rec}$).

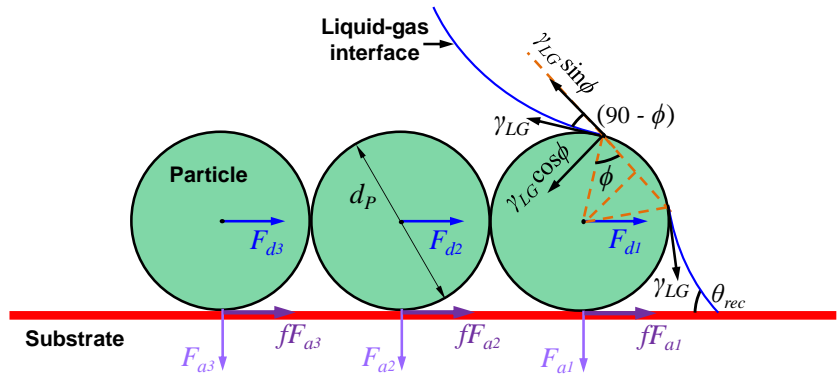

Figure 2: Schematic of the forces acting on the colloidal particles near the contact line. Adapted from Ref. [13], Copyright 2017, Springer Nature.

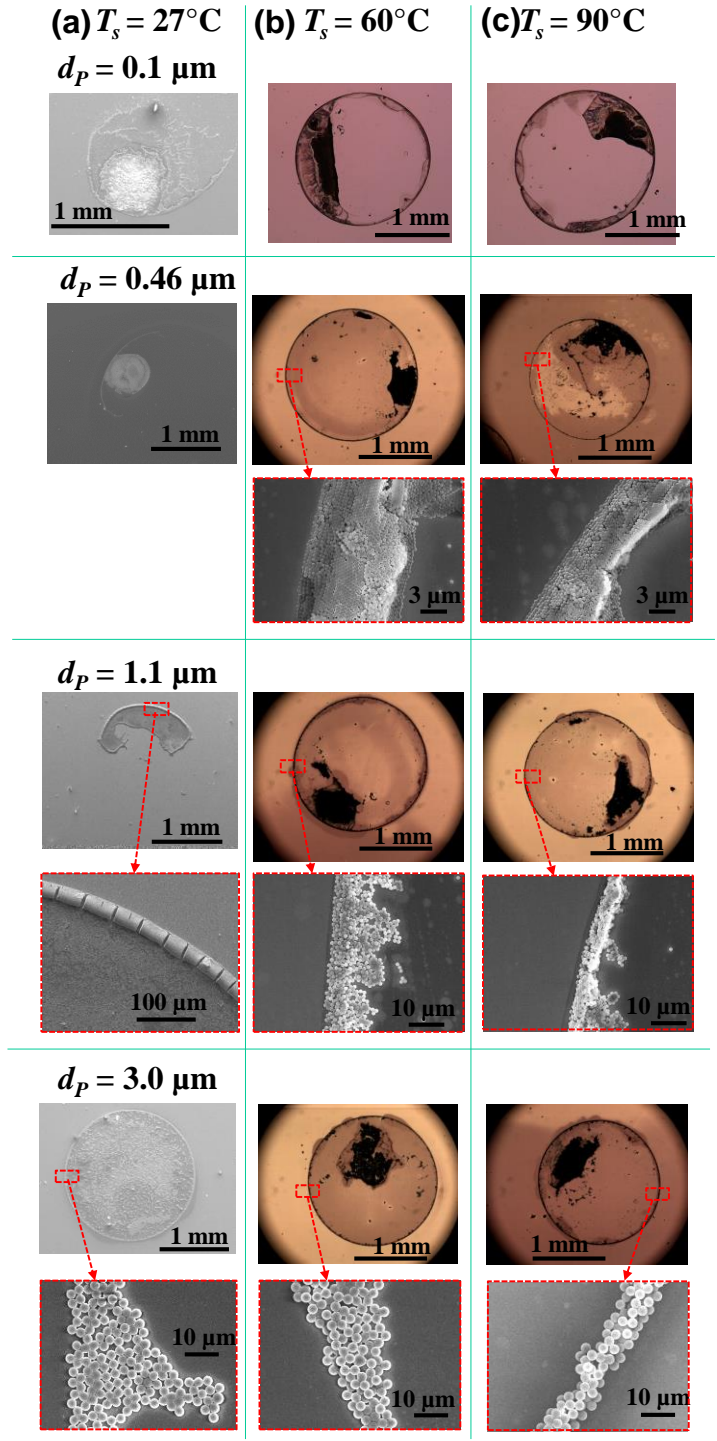

Figure 3: Effect of particle diameter and substrate temperature on deposit pattern formed after evaporation of a sessile aqueous droplet containing mono-dispersed colloidal polystyrene particles on wet-oxidized silicon. The particles concentration is 0.05% (v/v). Insets show a zoomed-in view of a part of the ring. Measured profiles of the ring for each case are given in supporting information.

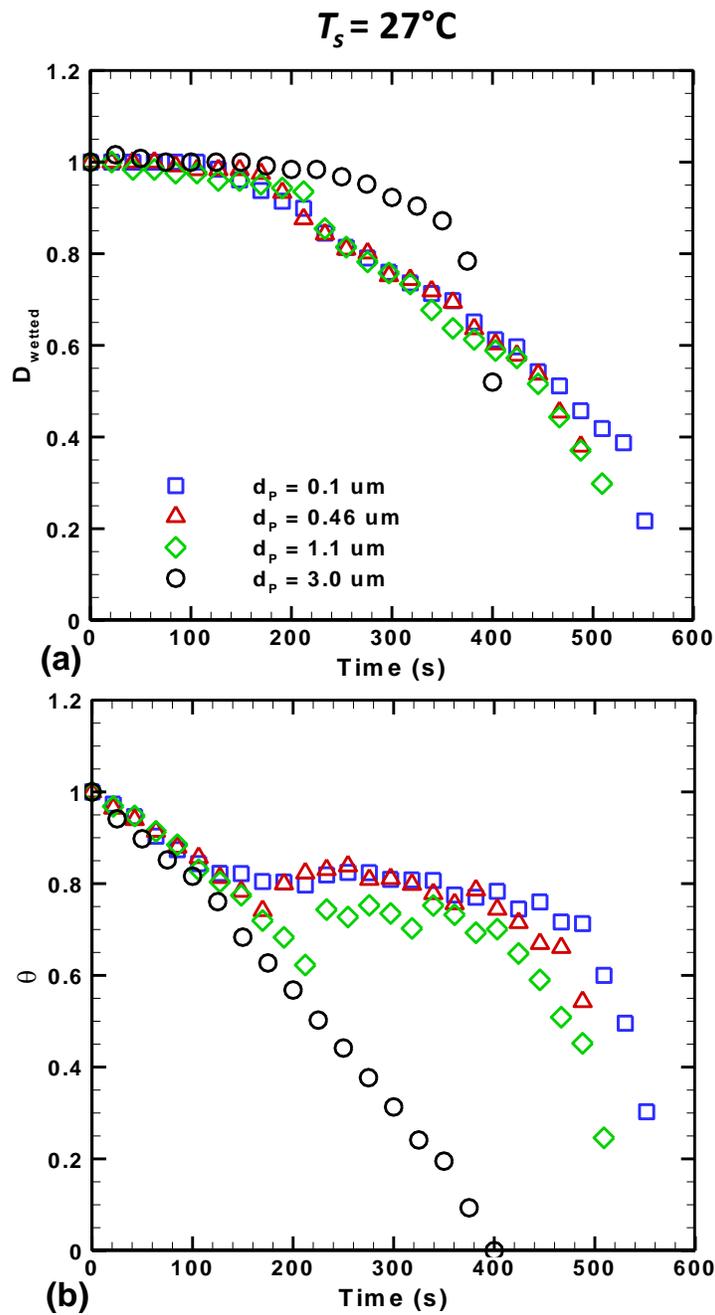

Figure 4: Time-variation of (a) droplet wetted-diameter ($D_{wetted}$) and (b) contact angle ($\theta$) during the evaporation of droplets containing mono-dispersed colloidal particles of different diameter on the nonheated silicon surface ($T_s = 27°C$).

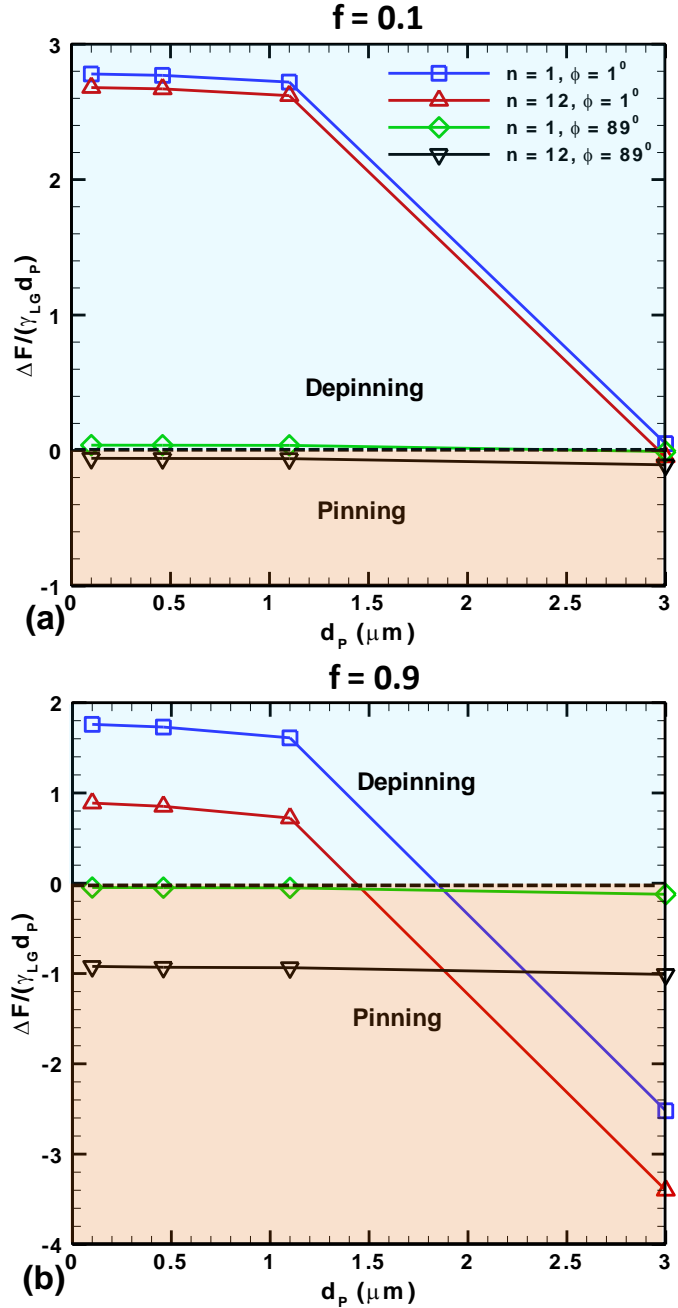

Figure 5: Variation of non-dimensional resultant force in a horizontal direction ($\Delta F^*$, eq. 7) as a function of particles diameter ($d_P$) for pinning and depinning of particles near the contact line during evaporation of droplets containing mono-dispersed colloidal particles. The sensitivity of the model is tested for a wide range of $\phi$ (1 and 89°), and friction factor (a) $f = 0.1$ and (b) $f = 0.9$. The values of the parameter used for calculations are given in the supporting information.

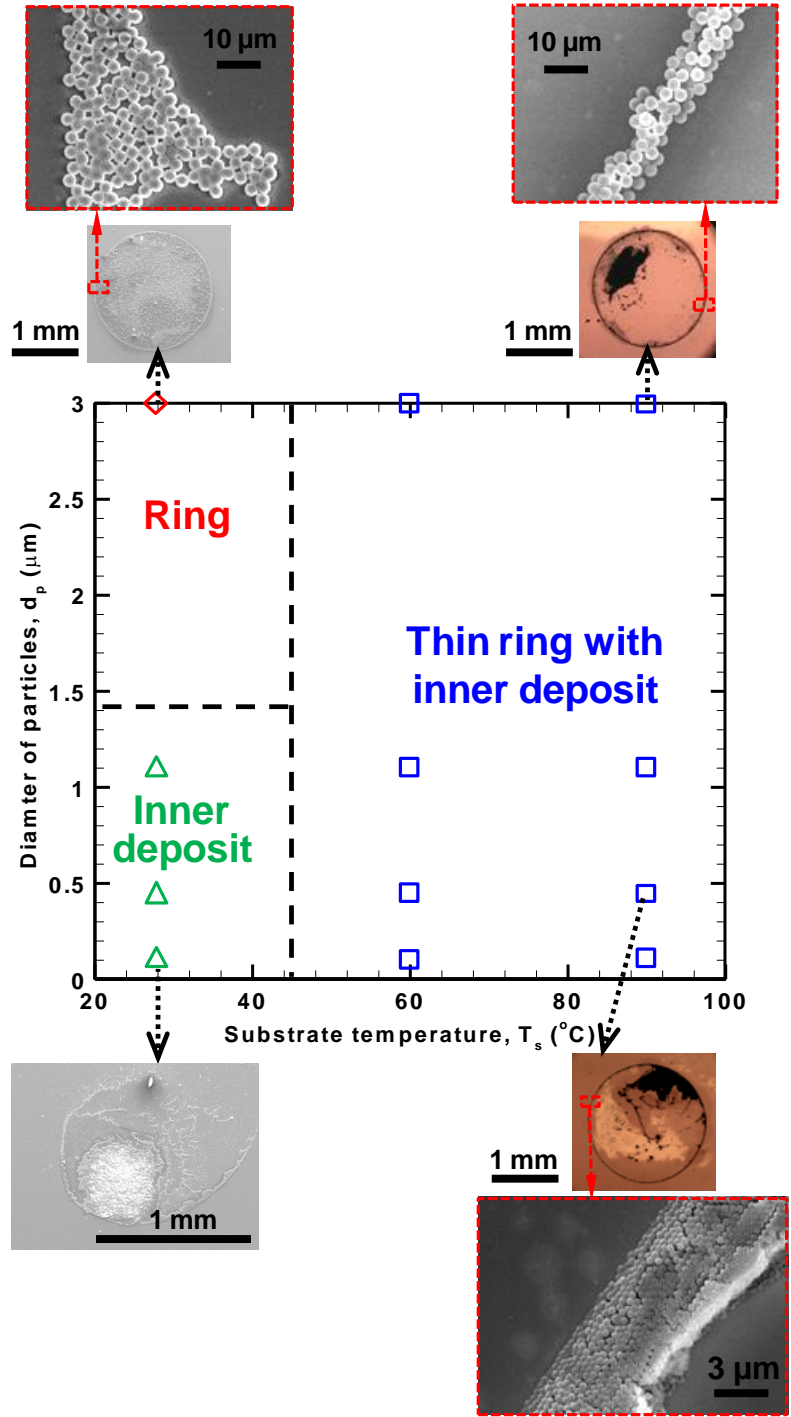

Figure 6: Regime map for predicting the deposit pattern as a function of the substrate temperature and diameter of particles. The dashed lines are plotted to demarcate the three regimes, inner deposit, ring and thin ring with inner deposit. Insets show the deposit image obtained by optical and scanning electron microscopy at particles concentration, $c = 0.05\%$. The insets are the representative images for the corresponding deposit pattern.

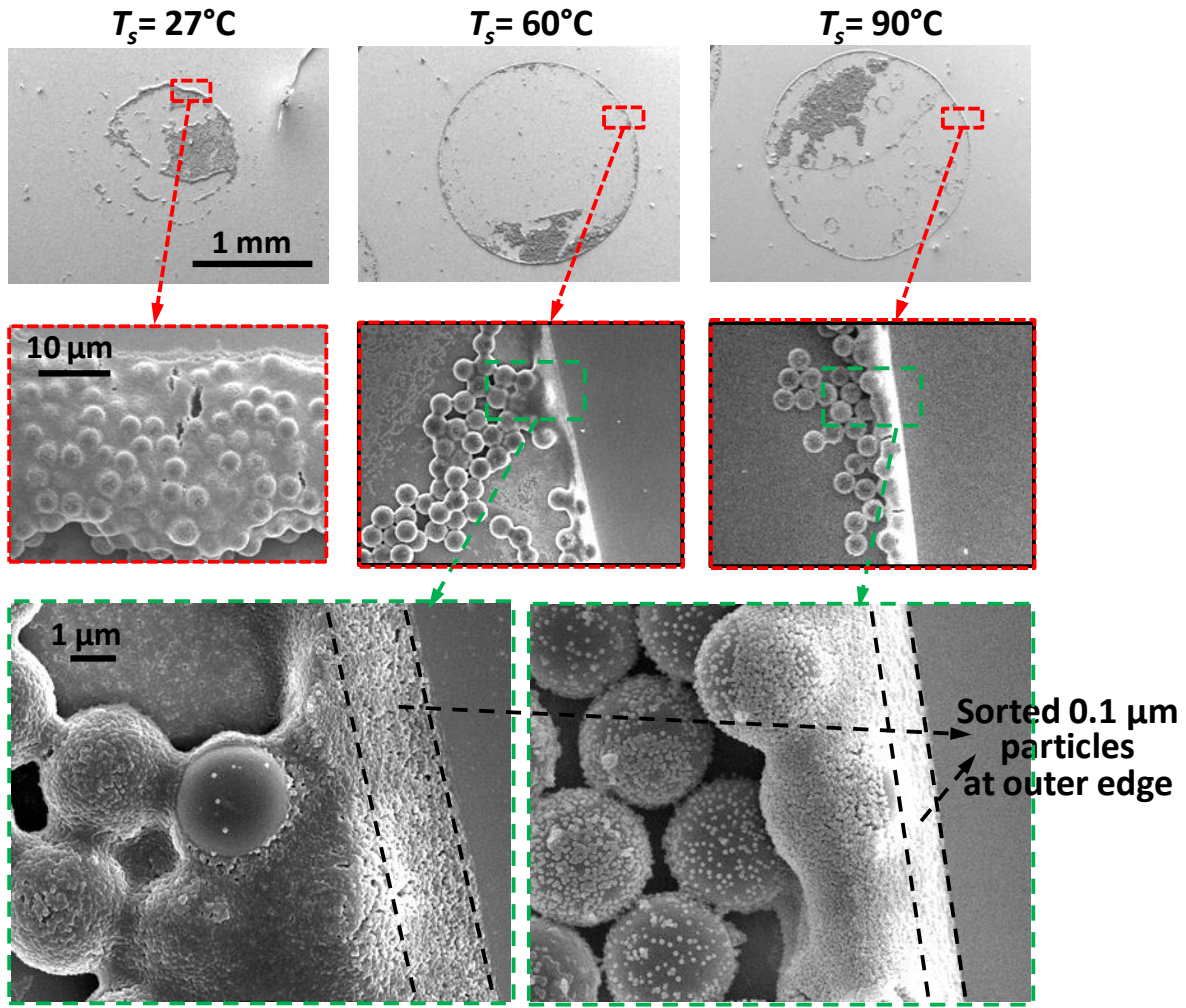

Figure 7: Effect of substrate temperature on the deposit pattern during sessile droplet evaporation containing bi-dispersed particles of diameter 0.1 and 3 μm (diameter ratio, $R$ = 0.03). Self-sorting occurs on heated substrate and smaller 0.1 μm particles deposit at the outer edge as compared to larger 3 μm particles.

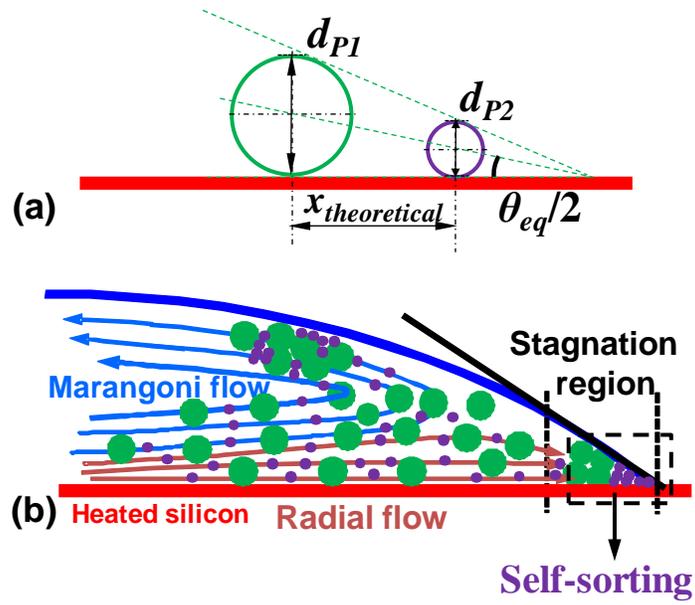

Figure 8: (a) Schematic of wedge assumption due to the geometry of contact line for deposition of bi-dispersed particles. (b) Mechanism of deposition of bi-dispersed colloidal particles at the contact line on the heated substrate.

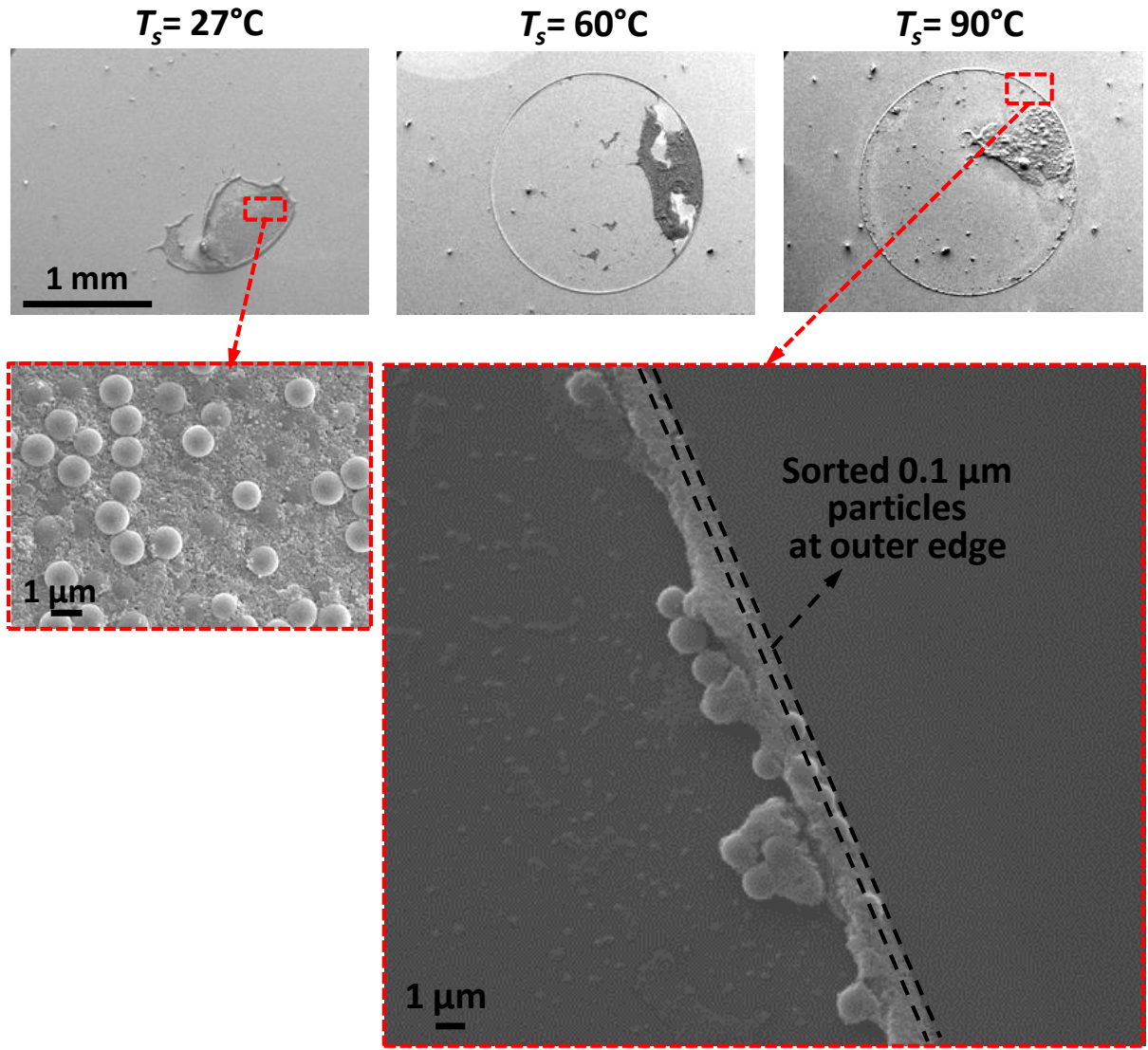

Figure 9: Effect of substrate temperature on deposit pattern during sessile droplet evaporation containing bi-dispersed particles of diameter 0.1 and 1.1 µm (diameter ratio, $R = 0.09$). Self-sorting occurs on heated substrate and smaller 0.1 µm particles deposit at the outer edge as compared to larger 1.1 µm particles.

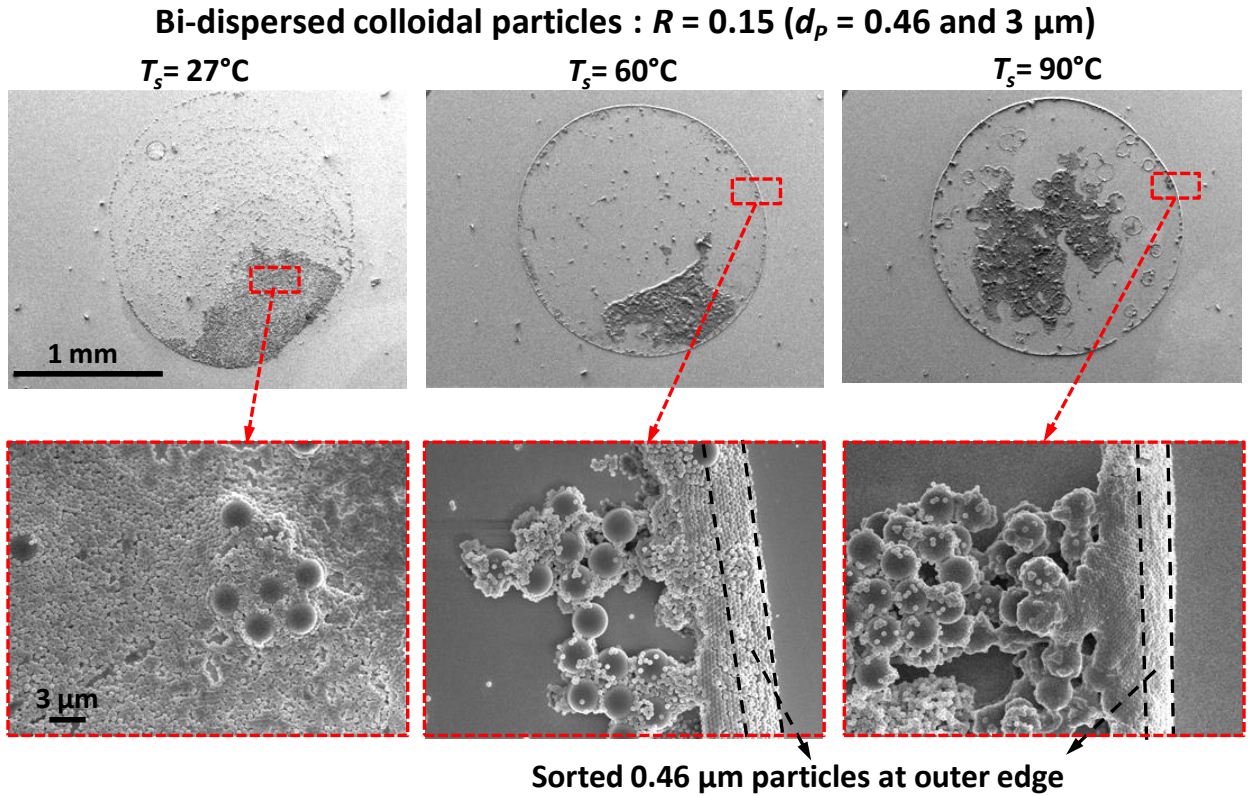

Figure 10: Effect of substrate temperature on deposit pattern during sessile droplet evaporation containing bi-dispersed particles of diameter - 0.46 and 3 μm (diameter ratio, $R = 0.15$). Self-sorting occurs on heated substrate and smaller 0.46 μm particles deposit at the outer edge as compared to larger 3 μm particles.

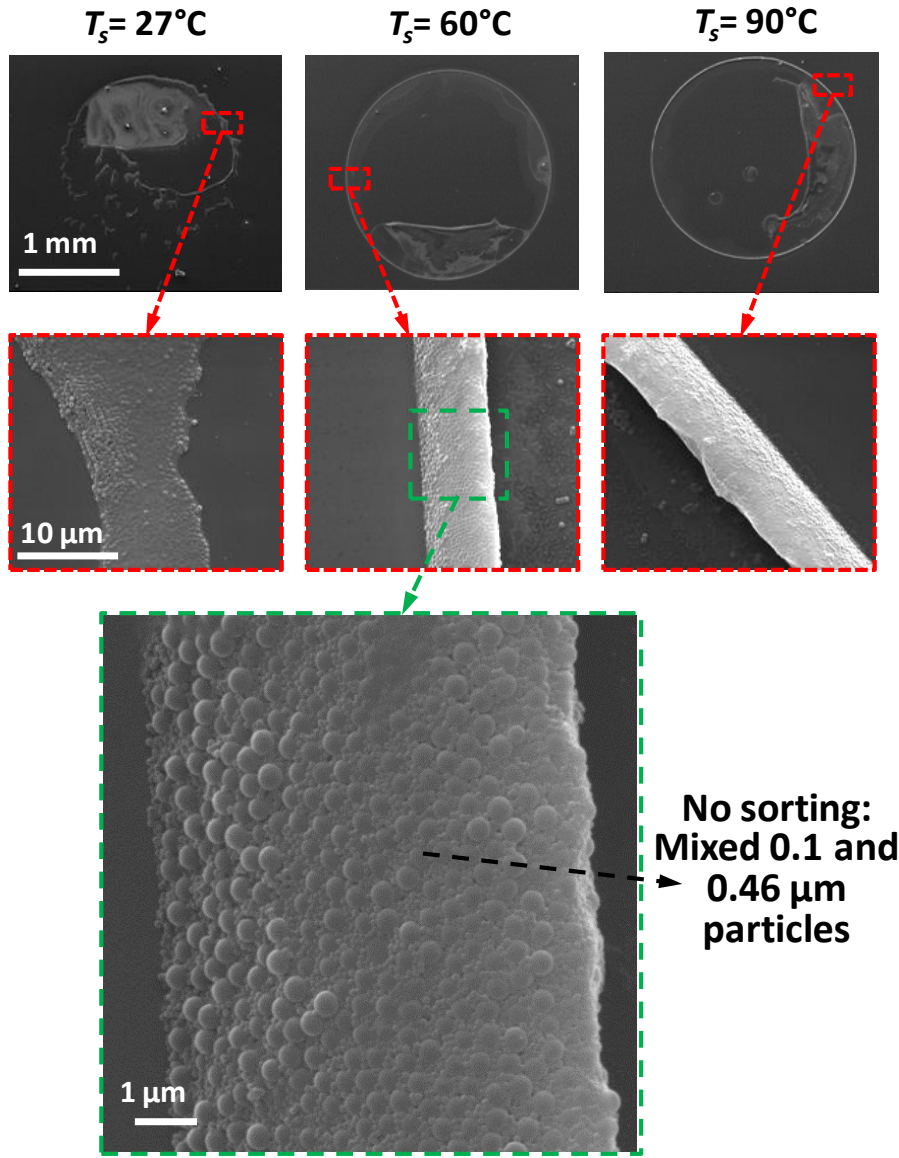

Figure 11: Effect of substrate temperature on deposit pattern during sessile droplet evaporation containing bi-dispersed particles of diameter 0.1 and 0.46 µm (diameter ratio, $R = 0.21$). Sorting does not occur and particles are almost uniformly mixed and get deposited at the contact line.

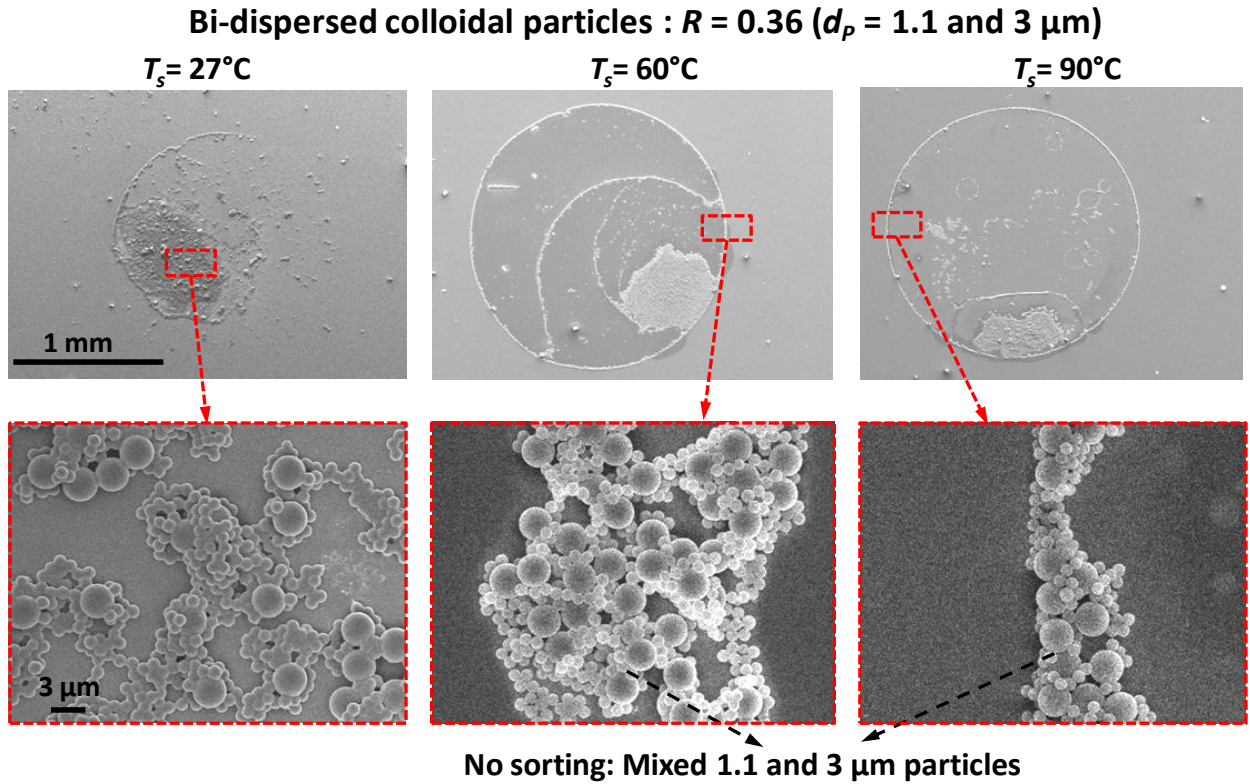

Figure 12: Effect of substrate temperature on deposit pattern and morphology formed after evaporation of a sessile droplet containing bi-dispersed particles of diameter 1.1 and 3 μm (diameter ratio, $R = 0.36$). Sorting does not occur and particles get almost uniformly mixed in the ring at $T_s = 60°C$ and $T_s = 90°C$.

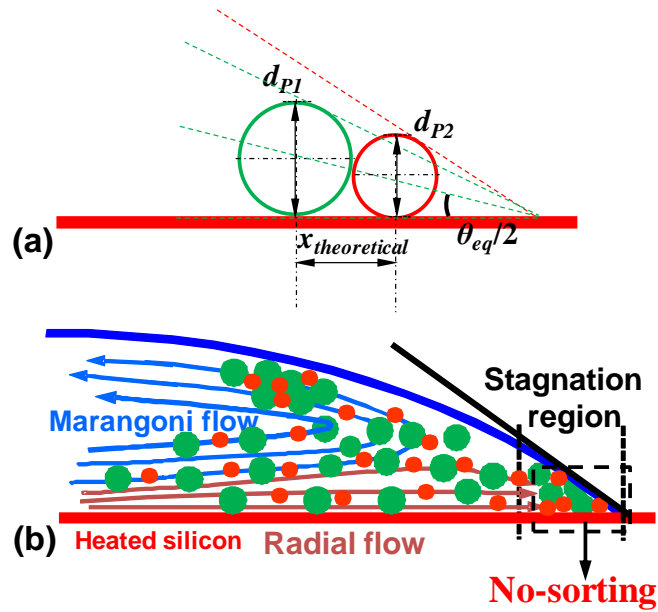

Figure 13: (a) Schematic of bi-dispersed particles of comparable diameter (or larger diameter ratio) near the contact line and (b) mechanism of their deposition.

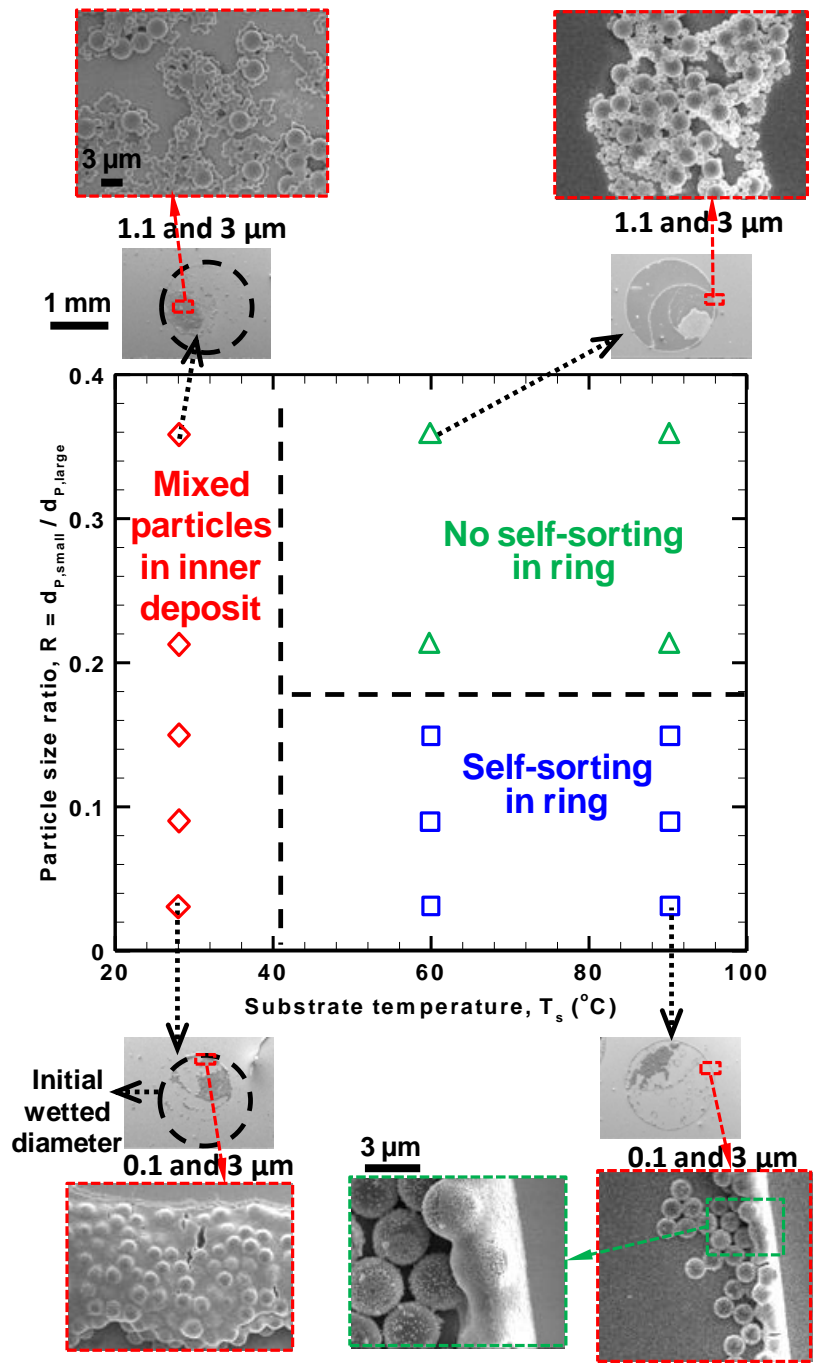

Figure 14: Regime map for the deposits of bi-dispersed colloidal particles as a function of substrate temperature and particle diameter ratio. Particles concentration is 0.05 % (v/v) in all cases. Dashed lines are plotted to demarcate the three regimes, mixed particles in inner deposit, self-sorting in the ring and no self-sorting in the ring. SEM images in insets show the morphology of the deposit. The insets are representative images for corresponding deposit pattern.